\documentclass[12pt]{article}
\usepackage{epsfig}
\usepackage{pst-plot,url,epsf}
\usepackage{axodraw}

\newcommand{\beq}{\begin{equation}}
\newcommand{\eeq}{\end{equation}}
\newcommand{\bea}{\begin{eqnarray}}
\newcommand{\eea}{\end{eqnarray}}

\newcommand{\epm}{e^+e^-}

\newcommand{\ra}{\rightarrow}
\def\earr{\end{array}}
\def\barr#1{\begin{array}{#1}}

\begin{document}
\thispagestyle{empty}
\begin{flushright}
October 2003\\
\vspace*{1.5cm}
\end{flushright}
\begin{center}
{\LARGE\bf Top quark pair production at a linear collider in the
           presence of\\[2mm] an anomalous $Wtb$ coupling}
\footnote{Presented by K.~Cie\'ckiewicz at the XXVII International Coference 
          of Theoretical Physics ``Matter to the Deepest: Recent Developments 
          in Physics of Fundamental Interactions'', Ustro\'n, Poland, 
          September 15--21, 2003.}$^,$~%
\footnote{Work supported in part by the Polish State Committee for Scientific 
        Research (KBN) under contract No. 2~P03B~045~23.}\\
\vspace*{2cm}
K. Cie\'ckiewicz\footnote{E-mail: cieckiew@server.phys.us.edu.pl}  
and K. Ko\l odziej\footnote{E-mail: 
         kolodzie@us.edu.pl}\\[1cm]
{\small\it
Institute of Physics, University of Silesia\\ 
ul. Uniwersytecka 4, PL-40007 Katowice, Poland}\\
\vspace*{2.5cm}
{\bf Abstract}\\
\end{center}
Angular distributions of a $\mu^+$ and a $b$-quark resulting from
the decay of a top quark 
produced at the $e^{+}e^{-}$ linear collider with
an unpolarized and a 100\% longitudinally polarized electron beam are 
presented. The results of the standard model are compared with
the results obtained in the presence of the anomalous $Wtb$ coupling.

\vfill
\newpage

\section{Introduction.}
Future high luminosity $e^{+}e^{-}$ linear collider with its very
clean experimental environment will be the most suitable tool
for searching for the effects of physics beyond the standard model (SM).
In particular, such effects may manifest themselves in deviations
of the top quark properties and interactions from those predicted by SM.
Therefore, experimental studies of the top quark pair
production belong to the research program of any future linear
collider \cite{NLC}. Due to its large decay width, the top quark decays even
before it hadronizes. As the dominant top decay mode is
\bea
\label{tbw}
t \ra bW^+,
\eea
it is interesting to look at extensions of the pure left-handed 
$Wtb$ coupling that governs reaction (\ref{tbw}) in
the lowest order of SM. 

Such extensions of SM can be best parametrized in terms of the effective
lagrangian that has been written down in Eq.~(3) of \cite{CK}. The 
corresponding modification of the SM Feynman rule for the $Wtb$-vertex 
is the following
\bea
\label{Wtb}
\barr{l}
\begin{picture}(90,80)(-50,-36)
\Text(-45,5)[lb]{$W_{\mu}^+, p$}
\Text(35,27)[rb]{$t, p_{t}$}
\Text(35,-27)[rt]{$\bar{b}, p_{\bar{b}}$}
\Vertex(0,0){2}
\ArrowLine(0,0)(35,25)
\ArrowLine(35,-25)(0,0)
\Photon(0,0)(-45,0){2}{3.5}
\end{picture} 

\earr
\barr{l}
\\[6mm]
\longrightarrow \quad
\Gamma^{\mu}_{Wtb}=-{g\over\sqrt{2}}V_{tb}\;
\left[\,\gamma^{\mu}\left(f_1^- P_- +f_1^+ P_+\right) \right.
                                                                   \\[3mm]
\qquad \qquad \qquad \qquad \left.-{{i\sigma^{\mu\nu}p_{\nu}}\over m_W}
\left(f_2^- P_- +f_2^+ P_+\right)\,\right].
\earr
\eea
In Eq.~(\ref{Wtb}), $V_{tb}$ is the element of
the Cabibbo-Kobayashi-Maskawa matrix,
$P_{\pm}=(1\pm\gamma_5)/2$ are chirality projectors, $p$ is the four 
momentum of the incoming $W^+$ and $f_{i}^{\pm}, \; i=1,2$, are the $Wtb$
vertex form factors. The SM vertex is reproduced with 
$f_{1}^{-}=1$ and $f_{1}^{+}=f_{2}^{-}=f_{2}^{+}=0$.

As the experimental value of $\left|V_{tb}\right|$ is $0.9990$--$0.9993$ 
\cite{PDG} and deviation 
of the (V+A) coupling $f_{1}^{+}$ from zero is severely constrained by 
the CLEO data on $b\rightarrow s\gamma$ \cite{cleo}, in the following
we set $V_{tb}=1$, $f_{1}^{-}=1$ and $f_{1}^{+}=0$ and consider
modifications of the $Wtb$ vertex by nonzero values of the other
two anomalous form factors $f_2^+$ and $f_2^-$ which are often referred to as
the magnetic type anomalous couplings. 
Typical values of the couplings $f_{2}^{\pm}$ discussed in this talk 
are \cite{peccei}
\begin{eqnarray}
\left|f_{2}^{\pm}\right| \sim {\sqrt{m_{b}m_{t}}\over {v}}\sim 0.1.
\end{eqnarray}
They contradict neither the unitarity limit obtained from the 
$t{\bar t}$ scattering at the TeV energy scale that gives the constraint 
$\left|f_{2}^{\pm}\right|$ $\leq0.6$~\cite{renard}, nor the limits 
that are expected  from the upgraded Tevatron, which
are of order $0.2$.

In practice, the measurement of the form factors of Eq.~(\ref{Wtb}) through
decay (\ref{tbw}) does not take place in the rest frame 
of the top quark. At the linear collider, the top quark pair is produced
in the process
\bea
\label{eett}
         e^+e^- \rightarrow  t \bar{t}
\eea
and, as $t$ and $\bar{t}$ almost immediately
decay into 3 fermions each, what one actually observes are reactions of
the form
\bea
\label{ee6f}
\epm \ra 6\,{\rm f},
\eea
where 6f denotes a 6 fermion  final state that is possible in SM.
Reactions (\ref{ee6f}) receive contributions typically from several
hundred Feynman diagrams, whereas there are only two $\gamma$ and $Z$ exchange
signal diagrams in the annihilation channel that contribute to (\ref{eett}).
Fortunately, the signal diagrams dominate the cross section over a wide
range of the centre of mass system (CMS) energies, see \cite{KK}. This
justifies the simplified approach used in this talk, in which 
only the two signal diagrams that contribute to (\ref{ee6f}) are kept
and all other, non-doubly resonant diagrams, are neglected.
In the next section, we present our 
numerical results for one specific semileptonic channel of (\ref{ee6f})
\bea
\label{eett6f}
 e^+e^- \rightarrow  t^* \bar{t}^* \rightarrow b \nu_{\mu} \mu^+ 
\bar{b} d \bar{u}
\eea
in the double resonance approximation for the $t$ and $\bar{t}$.
In particular, we address the issue of determining the
spin of the top-quark produced in reaction (\ref{eett6f}) by measuring
the angular distribution of the $\mu^+$ resulting from its
decay, the method first proposed in \cite{Jezabek}.

\section{Numerical results}

In this section, we present numerical results on the angular
distributions of the $b$-quark and $\mu^+$ of reaction (\ref{eett6f}) at
$\sqrt{s}=360$~GeV and $\sqrt{s}=500$~GeV, typical for a future 
linear collider.

The matrix elements corresponding to Eq.~(\ref{Wtb}) have been
programmed with the helicity amplitude method of \cite{KZ} 
and \cite{KJ} and then implemented into {\tt eett6f}, a Monte Carlo program 
for top quark pair production and decay into 6 fermions at linear 
colliders \cite{eett6f}. The calculation has been performed in the
fixed width scheme with the top quark mass 
$m_t=174.3$~GeV and the 3 body top quark decay width calculated to 
lowest order of the perturbation series, taking into account the modified 
$Wtb$ coupling given by Eq.~(\ref{Wtb}) and assuming $\cal{CP}$ conservation. 
Other physical parameters used in the calculation
have been taken from \cite{PDG}.

\begin{figure}[ht!]
\label{fig1}
\begin{center}
\setlength{\unitlength}{1mm}
\begin{picture}(30,30)(53,-50)
\rput(5.3,-6){\scalebox{0.55 0.55}{\epsfbox{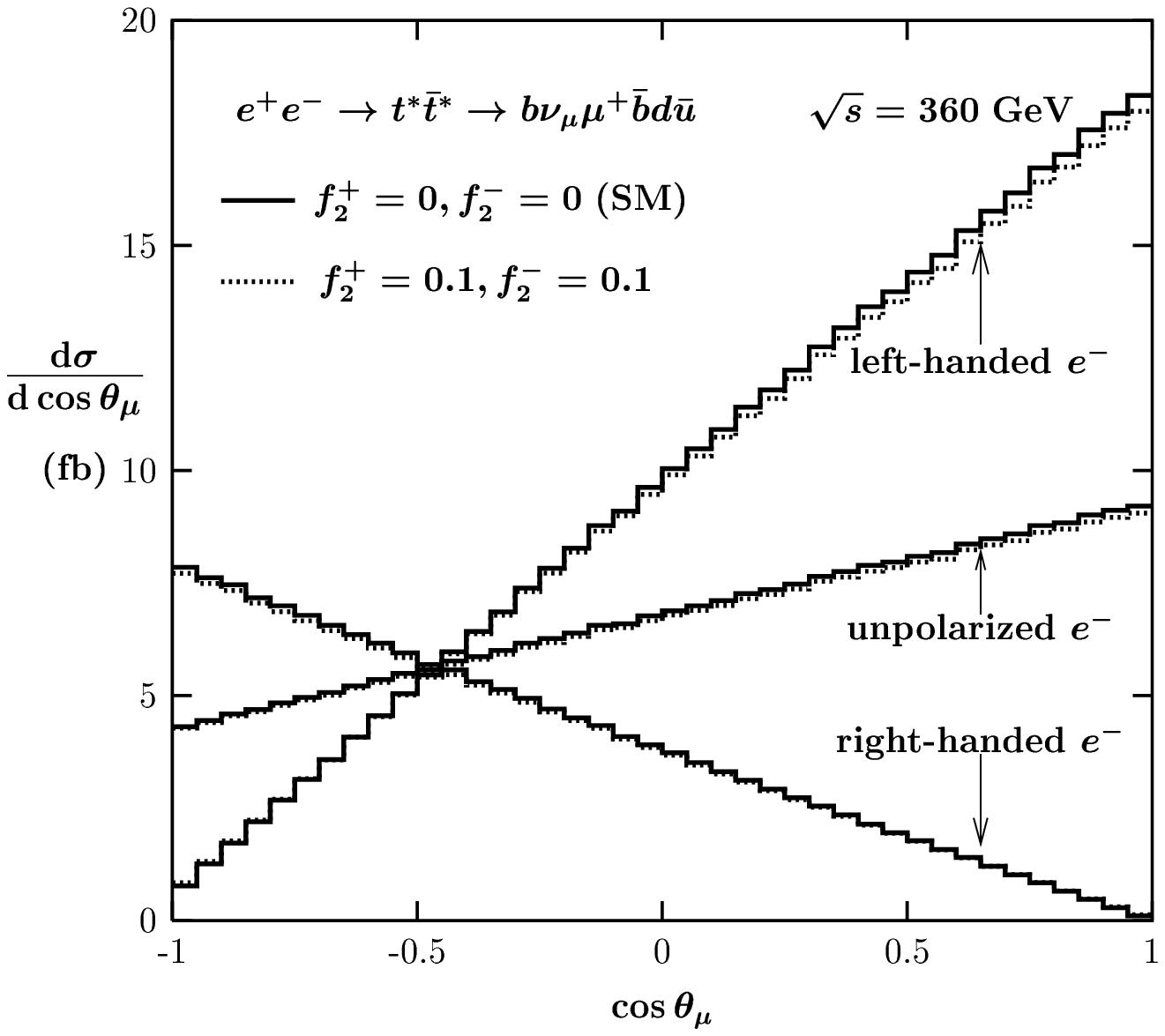}}}
\end{picture}
\begin{picture}(30,30)(17,-50)
\rput(5.3,-6){\scalebox{0.55 0.55}{\epsfbox{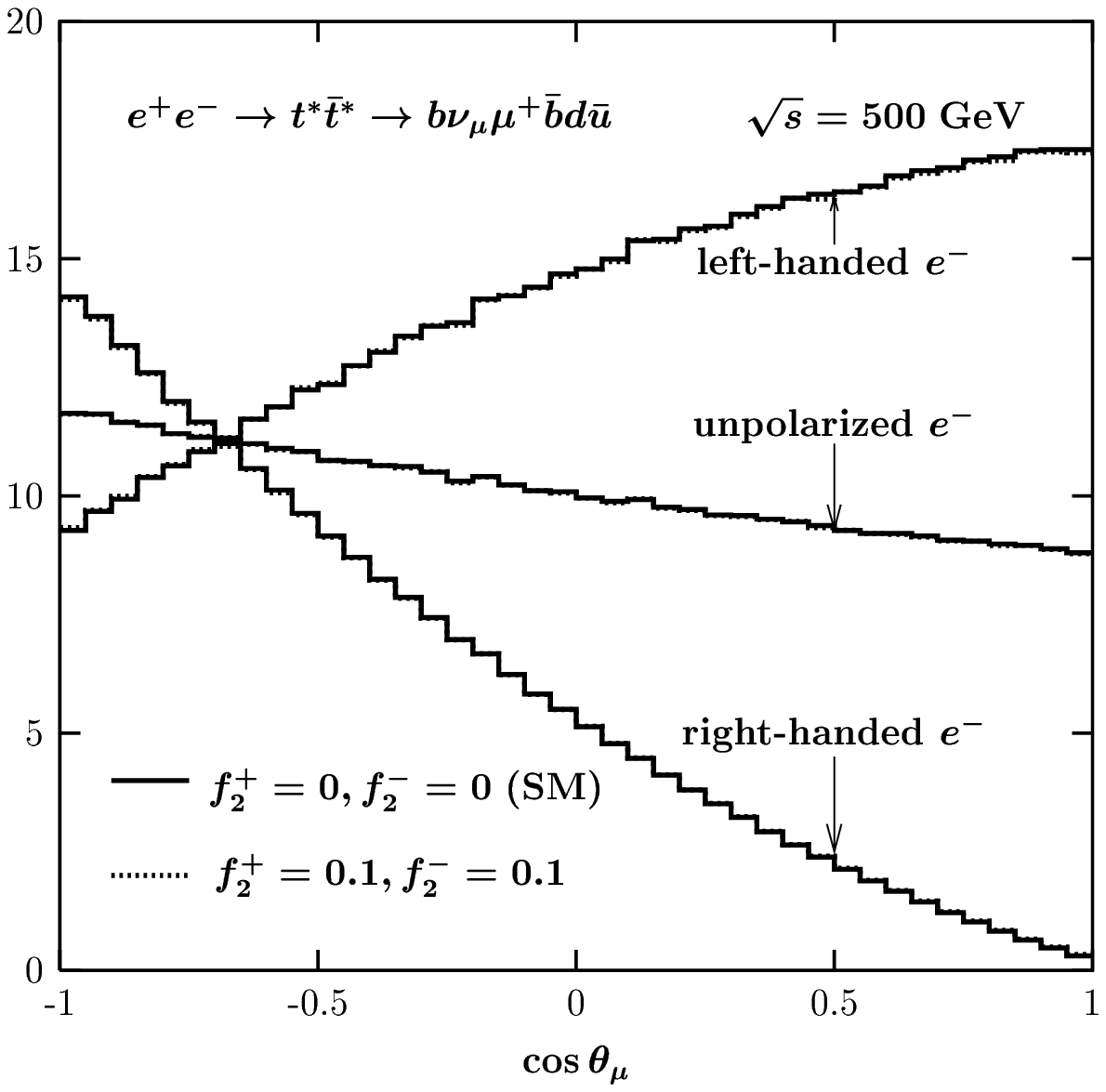}}}
\end{picture}
\end{center}
\vspace*{3.3cm}
\caption{Angular distributions of a $\mu^+$ at $\sqrt{s}=360$ GeV (left)
         and $\sqrt{s}=500$ GeV (right).}
\end{figure}

In Fig.~1, we plot the angular distributions of a $\mu^+$ at 
$\sqrt{s}=360$~GeV (left) and $\sqrt{s}=500$~GeV (right). The
plots show the differential cross section 
${\rm d}\sigma/{\rm d}\cos\theta_{\mu}$
for the unpolarized and for 100\% longitudinally polarized 
left-handed and right-handed electron beam.
The slant of the histograms representing unpolarized cross section 
both at $\sqrt{s}=360$~GeV and $\sqrt{s}=500$~GeV is 
caused solely by the
Lorentz boost of the corresponding flat angular distribution
of the $\mu^+$ resulting from the decay of unpolarized top
quark at rest. The slants of the histograms representing 
polarized cross sections at $\sqrt{s}=360$~GeV, on the other hand, reflect 
proportionality of the angular distribution 
of $\mu^+$ to $(1 + \cos\theta)$, if the spin of the decaying top-quark 
points in the positive direction of the $z$ axis (spin up),
and to $(1 - \cos\theta)$, if the spin of the decaying top-quark 
points in the negative direction of the $z$ axis (spin down),
see \cite{KK1} for illustration. With 100\% left-handedly (right-handedly) 
polarized electron beam that goes in the direction of negative $z$-axis, 
the top quark is produced preferably with the spin up (down). The corresponding
$(1 \pm \cos\theta)$ behaviour of the $\mu^+$ angular distribution
is somewhat changed by the Lorentz boost, in particular at
$\sqrt{s}=500$~GeV.
The dotted histograms in Fig.~1 represent the angular distributions 
of $\mu^+$ in the presence of anomalous $Wtb$ coupling (\ref{Wtb})
with $f_1^{\pm}$ set to their SM values, $f_1^+=0$, $f_1^-=1$, and 
$f_2^+=f_2^-=0.1$. Except for a rather small effect 
in case of the left-handed electron beam at $\sqrt{s}=360$~GeV, the change 
in the $\mu^+$ angular distributions is hardly visible. This
nicely confirms the decoupling theorem discussed in \cite{BHR}.

The angular distributions of a $b$-quark at $\sqrt{s}=360$~GeV and 
$\sqrt{s}=500$~GeV are plotted in Fig.~2. 
Again the dotted histograms represent the angular distributions 
of a $b$-quark in the presence of anomalous $Wtb$ coupling (\ref{Wtb})
with $f_1^{\pm}$ set to their SM values, $f_1^+=0$, $f_1^-=1$, and 
$f_2^+=f_2^-=0.1$. The numerical effect of the anomalous coupling
is bigger than in Fig.~1. It is visible in particular for the
longitudinally polarized electron beams.

\begin{figure}[ht!]
\label{fig2}
\begin{center}
\setlength{\unitlength}{1mm}
\begin{picture}(30,30)(53,-50)
\rput(5.3,-6){\scalebox{0.55 0.55}{\epsfbox{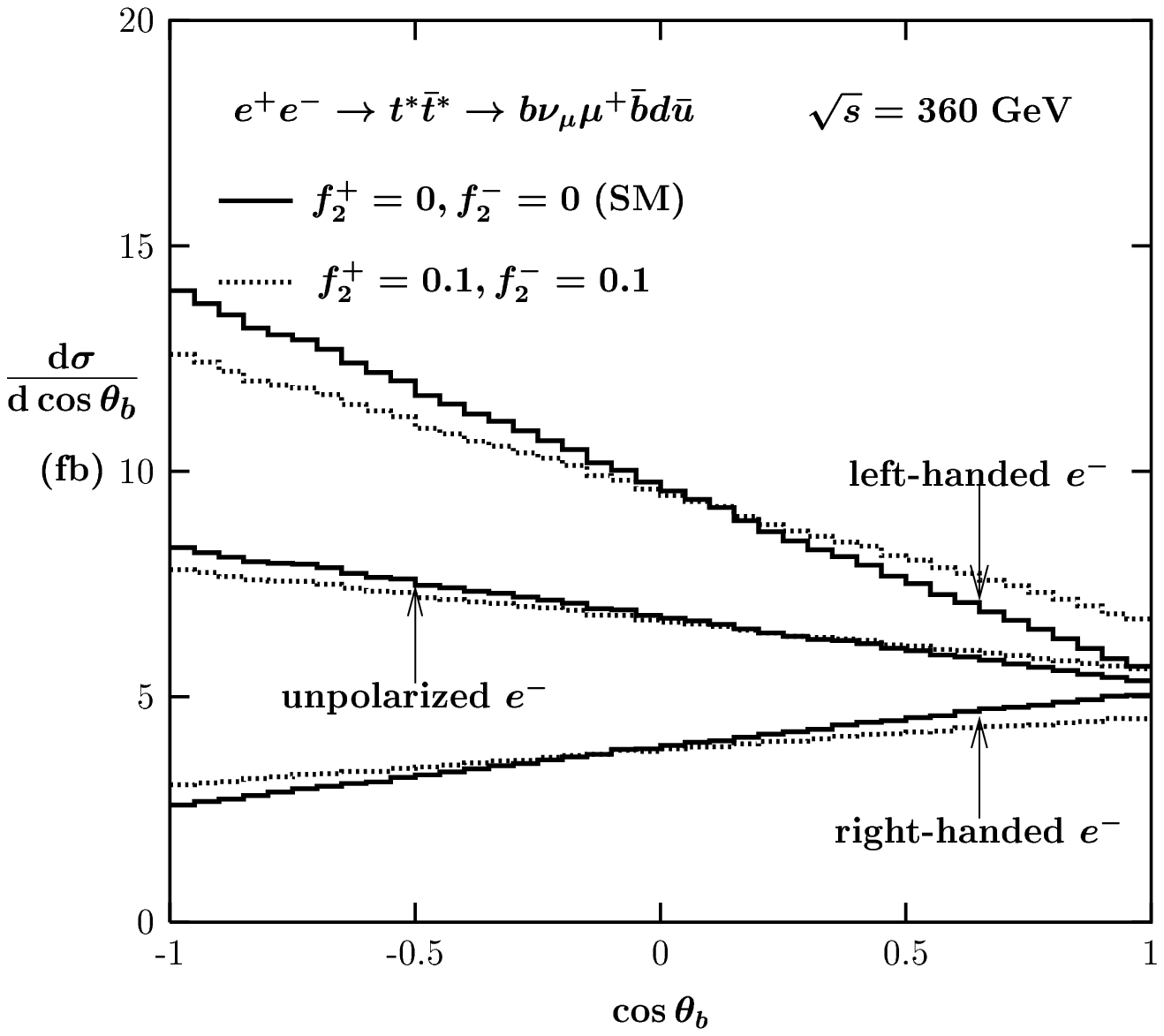}}}
\end{picture}
\begin{picture}(30,30)(17,-50)
\rput(5.3,-6){\scalebox{0.55 0.55}{\epsfbox{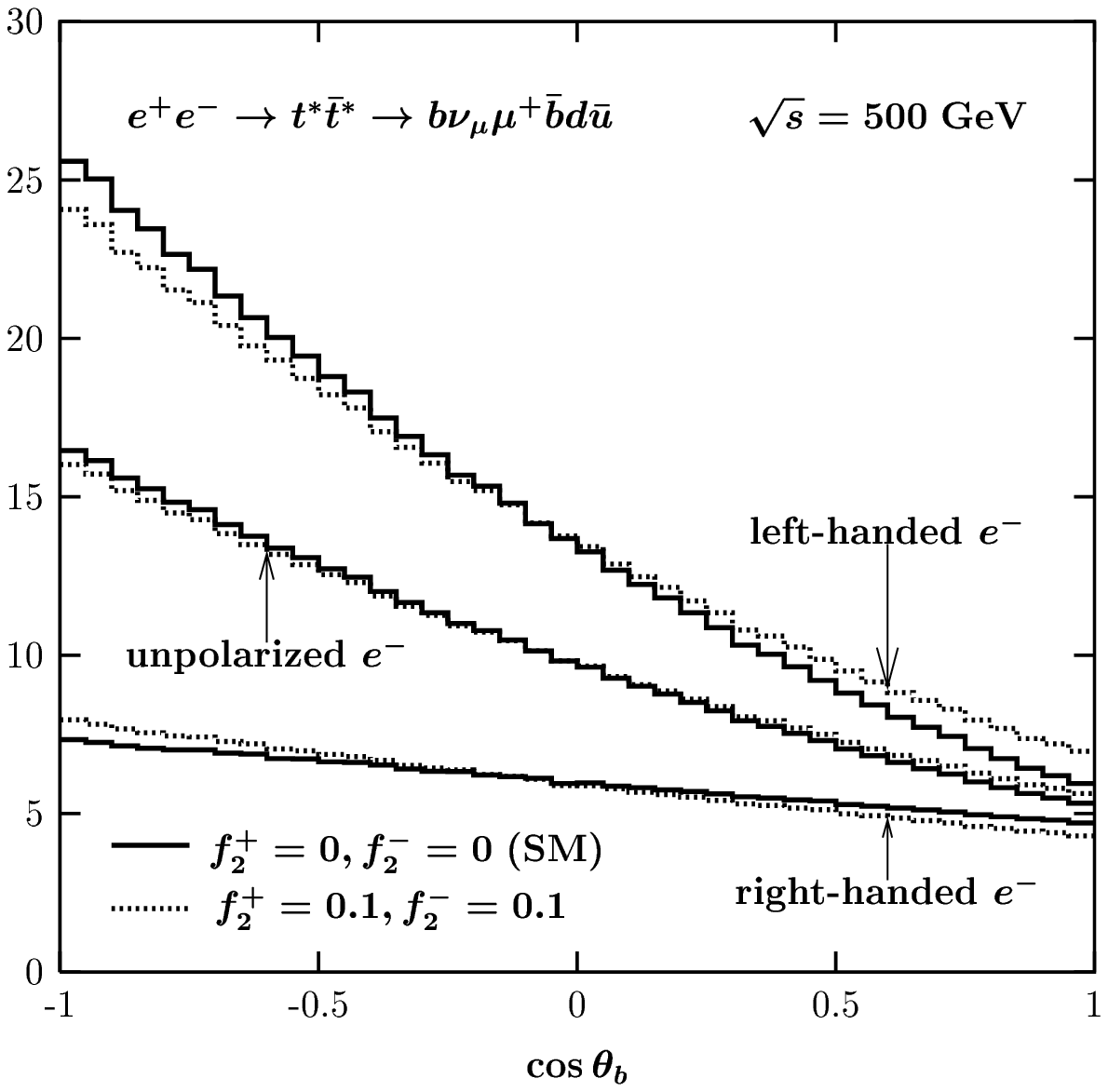}}}
\end{picture}
\end{center}
\vspace*{3.3cm}
\caption{Angular distributions of a $b$-quark at $\sqrt{s}=360$ GeV (left)
and $\sqrt{s}=500$~GeV (right).}
\end{figure}

\section{Summary}

We have computed angular distributions of a $\mu^+$ and a $b$-quark 
resulting from the decay of a top quark produced in reaction (\ref{eett6f})
at a linear collider. The results have been obtained 
with the unpolarized and 100\% longitudinally polarized electron beam.
Analysis of the $\mu^+$ distributions obtained with the longitudinally 
polarized beam shows that they are a very sensitive probe of the top quark
polarization, as expected. We have also illustrated 
how the anomalous $Wtb$ coupling modifies 
the $b$-quark angular distributions 
while it practically does not affect 
the angular distributions of $\mu^+$.

\begin{center}
{\bf Acknowledgement}
\end{center}

We would like to thank Z. Hioki and B. Grz\c adkowski for valuable remarks.

\end{document}